\documentclass[conference]{IEEEtran}

\usepackage[letterpaper, left=0.625in,right=0.625in,top=0.75in,bottom=1in]{geometry}

\usepackage{lipsum}

\usepackage{mathtools}

\usepackage{floatflt}

\usepackage{color}
\usepackage{cite}

\usepackage{graphicx}
\usepackage{amsmath}
\usepackage{amsthm}
\usepackage{amssymb}

\usepackage{verbatim}

\usepackage{psfrag,array}

\usepackage{subfigure}

\usepackage{stfloats}

\usepackage{todonotes}

\usepackage{amsmath}
\usepackage{epstopdf}
\usepackage{algorithmic}
\usepackage{algorithm}

\newcommand{\p}[1]{\mathop{\mbox{\it p} } }

\newcommand{\be}{\begin{equation}}
\newcommand{\ee}{\end{equation}}
\newcommand{\ba}{\begin{array}}
\newcommand{\ea}{\end{array}}
\newcommand{\bea}{\begin{eqnarray}}
\newcommand{\eea}{\end{eqnarray}}
\newcommand{\bean}{\begin{eqnarray*}}
\newcommand{\eean}{\end{eqnarray*}}

\newcommand{\rmh}{^{\rm H}}

\definecolor{white}{rgb}{1,1,1}

\newtheorem{property}{Property}
\newtheorem{definition}{Definition}

\usepackage{nopageno}

\begin{document}

\title{Capacity Degradation with Modeling Hardware Impairment in Large Intelligent Surface}
\author
{
Sha Hu, Fredrik Rusek, and Ove Edfors\\
Department of Electrical and Information Technology, \\Lund University, Lund, Sweden\\ \{firstname.lastname\}@eit.lth.se.}

\maketitle

\thispagestyle{empty}

\begin{abstract}
In this paper, we consider capacity degradations stemming from potential hardware impairments (HWI) of newly proposed Large Intelligent Surface (LIS) systems. Without HWI, the utility of surface-area (the first-order derivative of the capacity with respect to surface-area) is shown to be proportional to the inverse of it. With HWI, the capacity as well as the utility of surface-area are both degraded, due to a higher effective noise level caused by the HWI. After first modeling the HWI in a general form, we derive the effective noise density and the decrement of utility in closed-forms. With those the impacts of increasing the surface-area can be clearly seen. One interesting but also natural outcome is that both the capacity and utility can be decreased when increasing the surface-area in the cases with severe HWI. The turning points where the capacity and the utility start to decrease with HWI can be evaluated from the derived formulas for them. Further, we also consider distributed implementations of a LIS system by splitting it into multiple small LIS-Units, where the impacts of HWI can be significantly suppressed due to a smaller surface-area of each unit.
\end{abstract}

\section{Introduction}

Large Intelligent Surface (LIS) is a newly proposed wireless communication system \cite{HRE171, HRE172} that is beyond massive MIMO \cite{M10, MM12, MM14} and breaks the traditional antenna-array concept. As envisioned in \cite{HRE171, HRE172}, a LIS allows for an unprecedented focusing of energy in three-dimensional space, remote sensing with extreme precision, and unprecedented data-transmissions, which fulfills visions for the 5G communication systems \cite{AZ14} and the concept of Internet of Things \cite{IoT}. The abundant signal dimensions of the received signal with a LIS system can also facilitate potential applications using artificial intelligence \cite{JH17}. 

In \cite{HRE171}, fundamental limits on the number of independent signal dimensions are derived under the assumption of a single deployed LIS with infinite surface-area. The results reveal that with matched-filter (MF) processing at the LIS, the inter-user interference of two users is close to a \textit{sinc}-function. Consequently, as long as the distance between two users are larger than half the wavelength, the inter-user interference is negligible \cite{HE18}. In \cite{HRE172}, fundamental limits on positioning with LIS are also derived, and the results show that the Cram\'er-Rao lower-bound (CRLB) for positioning can decrease in the third-order of the surface-area of the LIS. These nice properties show the potential of LIS in future wireless communication systems.

Implementing the LIS, on the other hand, is challenging and brings many new research questions. One potential issue is the hardware impairments (HWI) such as Tx-RF impairments \cite{JT17, SB10}, analog imperfectness and quantization errors \cite{BD14, ZS17}, nonlinearity of the power amplifies \cite{MN10}, time and frequency synchronization errors \cite{M04}, etc. These HWI are commonly encountered in current communication systems, but with LIS it gets more severe since the surface-area of LIS is typically large (for instances, using facades of buildings or long walls in airports as LIS). Hence, the HWI can degrade the capabilities of the LIS in practical implementations.

In this paper, we consider the HWI in analyzing the capacity of LIS, and target at understanding of interplays among the surface-area, the HWI, and the degradation of capacity and utility. We model the HWI in the form of errors where the center of a LIS is used as a reference point in hardware designs, and the impairments are caused by the distance from the considered point to the center on the LIS. Such a modeling implies that the farer from the center, the larger HWI will present in the signal processing unit (SPU), which collects the received signal reached at the LIS. The HWI are modeled by a non-negative function with using the distance $r$ of the consider point to the center of the LIS as a variable. With such a principle, we carry out analytical analysis of the capacity degradation with the HWI. We also measure the utility degradation of the surface-area, i.e., the slope of capacity in relation to the surface-area $\mathcal{A}$, which can be used as references when implementing a LIS system. That is, when the utility is below a certain threshold or even negative, it is not cost effective to further increase the surface-area. 
 
Further, as shown in \cite{HE18}, compared to a centralized deployment, a LIS system that comprises a number of small LIS-Units has several advantages. Firstly, the surface-area of each LIS-Unit can be sufficiently small which facilitates flexible deployments and configurations. Secondly, LIS units can be added, removed, or replaced without significantly affecting system design. A distributed deployment of LIS system is also beneficial to suppress the HWI, due to the fact that the surface-area of each LIS-Unit is smaller compared to a single LIS. We also extensively analyze the impacts of HWI in a LIS system that comprises an array of small LIS-Units, and show that the HWI can be significantly suppressed, with a higher cost of implementation complexity to build such an array.

\begin{figure*}[t]
\begin{center}
\vspace*{-4mm}
\hspace*{-5mm}
\scalebox{1.1}{\includegraphics{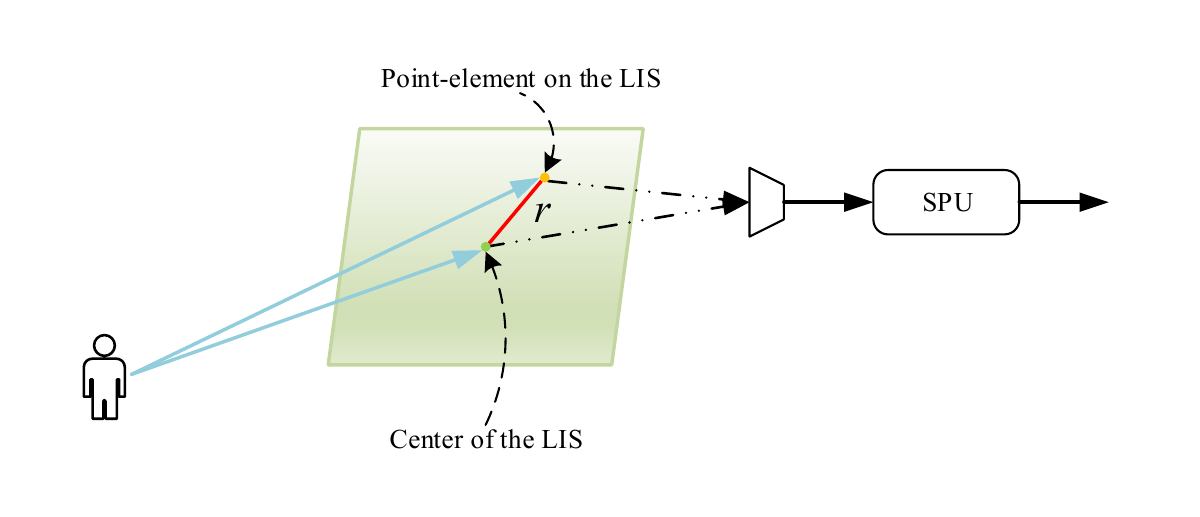}}
\vspace*{-10mm}
\caption{\label{fig1}An illustration of a possible implementation of the LIS system, where an SPU collects the radiation signal across the entire surface. The reference point is the center of the LIS, and possible HWI at other points increase with the distance to the center point.}
\vspace*{-6mm}
\end{center}
\end{figure*}

\section{Narrowband Received Signal Model with LIS}

We consider a transmission from a single-antenna user\footnote{Due to the strong interference-suppressing capability of the LIS, we consider the case of a single user. For multiple users, as they almost do not interfere each other \cite{HE18}, the analysis for each user remains the same.} located in a three-dimensional space to a two-dimensional LIS deployed on the plane $z\!=\!0$ as depicted in Fig. \ref{fig1}. Expressed in Cartesian coordinates, the center of the LIS\footnote{When referring to the coordinates of points on the LIS, the $z$-dimension is omitted as it is zero by default.} is located at $(0, 0)$, and the user is located at $(x_0,y_0, z_0)$, where $z_0\!>\!0$ and $x_0$, $y_0$ are arbitrary values. Under perfect line-of-sight (LoS) propagation, the user transmits a symbol $a$ to the LIS where $\mathbb{E}\{|a|^2\}\!=\!1$.

\subsection{Received Signal Model without HWI}

Denoting $\lambda$ as the wavelength and following \cite{HRE171, HRE172}, the effective channel $s_{x_0,\,y_0,\,z_0}(x,y)$ at position $(x, y)$ on the LIS can be modeled\footnote{Here we assume a narrowband system where the transmit time from the user to the LIS is negligible compared to its symbol period.} as
\bea \label{md} s(x,y)=\frac{1}{2}\sqrt{\frac{z_0}{\pi}}\eta^{-\frac{3}{4}}\exp\!\left(\!-\frac{2\pi j\sqrt{\eta}}{\lambda}\right)\!,   \eea
where the metric
\bea \eta=(x_0-x)^2+(y_0-y)^2+z_0^2,\;\, \text{for}\; (x,\,y)\in\mathcal{S}, \notag \eea
and $\mathcal{S}$ denotes the surface spanned by the LIS with an area $\mathcal{A}$.

Based on (\ref{md}), the received signal at $(x, y)$ of the LIS  is
\bea \label{rxyt} r(x,y) = \sqrt{P}s(x,y)a +n(x,y),\eea
where $P$ is the transmit power, and $n(x,y)$ is AWGN with a spatial power spectral density $N_0$ across the LIS. The average received energy stemming from a single symbol $a$ at the LIS, according to (\ref{rxyt}), equals
 \bea \label{g1} P\iint\limits_{(x,\,y)\in\mathcal{S}}|s(x,y)|^2 \mathrm{d}x\mathrm{d}y = \zeta P,\eea
where
\bea \label{zeta} \zeta= \frac{z_0}{4\pi}\iint\limits_{(x,\,y)\in\mathcal{S}}\eta^{-\frac{3}{2}} \mathrm{d}x\mathrm{d}y.\eea
Without loss of generality and in order to simplify the capacity analysis, we assume the user to be located on a central perpendicular line (CPL)\footnote{Such an ideal assumption holds when the LIS is sufficiently large or small. For other cases, approximation techniques in [2] can be applied to carry out the analysis based on the results obtained with the CPL case.}, that is $x_0\!=\!y_0\!=\!0$. Further, we assume a square\footnote{Similar analysis can be carried out for other shapes of the LIS, or alternatively, we can transfer the LIS of other shapes into a square one with the same surface-area.} LIS with $-A\!\leq \!x\!\leq\! A$ and $-A\!\leq\! y\!\leq\! A$. 

Then, it can be shown that $\zeta$ equals [1]
\bea \label{nu} \zeta\!\!\!\!&=&\!\!\!\!\frac{1}{\pi}\arctan\!\left(\frac{A^2}{z_0\sqrt{2A^2+z_0^2}} \right)\notag \\
\!\!\!\!&=&\!\!\!\!\frac{1}{\pi}\arctan\!\left(\frac{\tau^2}{\sqrt{2\tau^2+1}} \right)\!, \eea
where we defined a normalized length of the LIS (by the distance $z_0$) as
\bea \tau=A/z_0, \eea
and \lq\lq{}$\arctan$\rq\rq{}denotes the inverse \textit{tangent} function. This is the array gain obtained with the LIS for any finite deployed surface-area. If the LIS is infinitely large, i.e., $A\!=\!B\!=\!\infty$, then it holds that $\zeta\!=\!1/2$, which means half of the isotropically transmitted power from the user is received (which is the maximal power that can be received) by the LIS.

\subsection{Modeling the Hardware Impairments (HWI)}

Hardware impairments can result from many aspects in the transmit and receiver chains even with calibration and compensation techniques \cite{JH17, SW17}. With LIS we model the HWI (or remaining HWI after mitigations) as a Gaussian process, represented by a variable $h(r)$ in the received signal, where
$$r=\sqrt{x^2+y^2}.$$
With such an assumption, the received signal (\ref{rxyt}) changes to
\bea \label{r1} r(x,y) \!\!\!\!&=&\!\!\!\! \sqrt{P}\left(1+h(r)\right)s(x,y)a +n(x,y) \notag \\
\!\!\!\!&=&\!\!\!\!\sqrt{P}s(x,y)a +w(x,y),\eea
where the effective noise comprising HWI equals
$$ w(x,y)=\sqrt{P}h(r)s(x,y)a+n(x,y).$$

The zero-mean variable $h(r)$ is Gaussian and its variance is a non-negative function with respect to the distance $r$, which we denote as $f(r)$ and its expressions will be discussed later. The rationale behind (\ref{r1}) is that, we model the HWI at each point on the LIS as a function of the distance to the center. This can be due to that we use the central point of the LIS as a reference point in hardware design, and the larger the distance from the center, the more severe the HWI is (more difficult to synchronize and tune the hardwares).

Taking (\ref{r1}) into consideration, the received power at the LIS remains the same as (\ref{g1}), but the effective noise density (with MF process\footnote{Although with HWI, the MF process is not optimal anymore since the noise density now is not flat across the LIS, we still assume the same MF process and use it as an approximation to the capacity due to simplicity.}) changes to
\bea \label{tN} \tilde{N}= N_0+\frac{z_0P}{4\pi}\;\,\frac{\iint\limits_{(x,\,y)\in\mathcal{S}}f(r)\eta^{-3} \mathrm{d}x\mathrm{d}y}{\iint\limits_{(x,\,y)\in\mathcal{S}}\eta^{-\frac{3}{2}} \mathrm{d}x\mathrm{d}y}.\eea
The proof of (\ref{tN}) is given in Appendix. As the surface-area $\mathcal{A}$ increases, the effective noise density may get higher, therefore it is of interest to analyze the capacity changes with the HWI for guiding practical implementations of LIS systems.

\subsection{Capacity Bound with the LIS}
The capacity (nat/s/Hz) corresponding to the received model (\ref{r1}) equals \cite{HRE171}
\bea \label{C1} \mathcal{C}=\log\left(1+\frac{\zeta P}{\tilde{N}}\right)\!, \eea
where $\zeta$ is given in (\ref{nu}).

\begin{definition}
To analyze the changes of $\mathcal{C}$ in relation to the surface-area, we define a utility $\gamma$ (nat/s/Hz/m$^2$) of the LIS as
\bea \label{gam} \gamma \triangleq  \frac{\partial\mathcal{C}}{\partial \mathcal{A}}= \frac{1}{8A}\frac{\partial\mathcal{C}}{\partial A}. \eea
\end{definition}

From (\ref{C1}), it can be shown that
\bea \label{gam1} \gamma=\frac{P}{8A(\tilde{N}+\zeta P)}\left(\frac{\partial \zeta}{\partial A}-\frac{\zeta }{\tilde{N}}\frac{\partial \tilde{N}}{\partial A}\right)\!,\eea
where the first-order derivatives
\bea \label{dev1} \frac{\partial \zeta}{\partial A}= \frac{2\tau }{\pi z_0\sqrt{2 \tau^2 +1} \left(\tau^2 + 1\right)}\!,   \eea
and 
\bea \label{dev2} \frac{\partial \tilde{N}}{\partial A}\geq 0,  \eea
are yet to be discussed. In an ideal case when there is no HWI, we have $\tilde{N}\!=\!N_0$ and the equality in (\ref{dev2}) holds. In this case, the utility $\gamma$ is maximized.

To bound $\gamma$, from (\ref{gam1}) it holds that 
\bea \label{gam2} \gamma\!\!\!\!&\leq&\!\!\!\!\frac{P}{8A(\tilde{N}+\zeta P)}\frac{\partial \zeta}{\partial A}\notag \\ \!\!\!\!&\leq&\!\!\!\!\frac{1}{8A\zeta}\frac{\partial \zeta}{\partial A} \notag \\ 
\!\!\!\!&=&\!\!\!\!\gamma_0,\eea
where
\bea \label{ubd} \gamma_0=\frac{1}{4z_0^2}\frac{1}{\arctan\left(\frac{\tau^2}{\sqrt{2\tau^2+1}} \right)} \frac{1}{\sqrt{2 \tau^2 + 1} \left(\tau^2 + 1\right)}. \eea
As can be seen, the upper-bound in (\ref{gam2}) is tight under two conditions: 1) there is no HWI, and 2) $N_0$ is negligible, i.e., in high signal-to-noise ratio (SNR) cases. Further, for a given $\tau$, the bound of utility $\gamma_0$ decreases quadratically in the distance $z_0$. That is, when $z_0$ increases, the length $A$ must increase faster than $z_0$ in order to have the same utility of the surface-area.

\begin{figure}[t]
\begin{center}
\vspace*{-4mm}
\hspace*{-5mm}
\scalebox{0.37}{\includegraphics{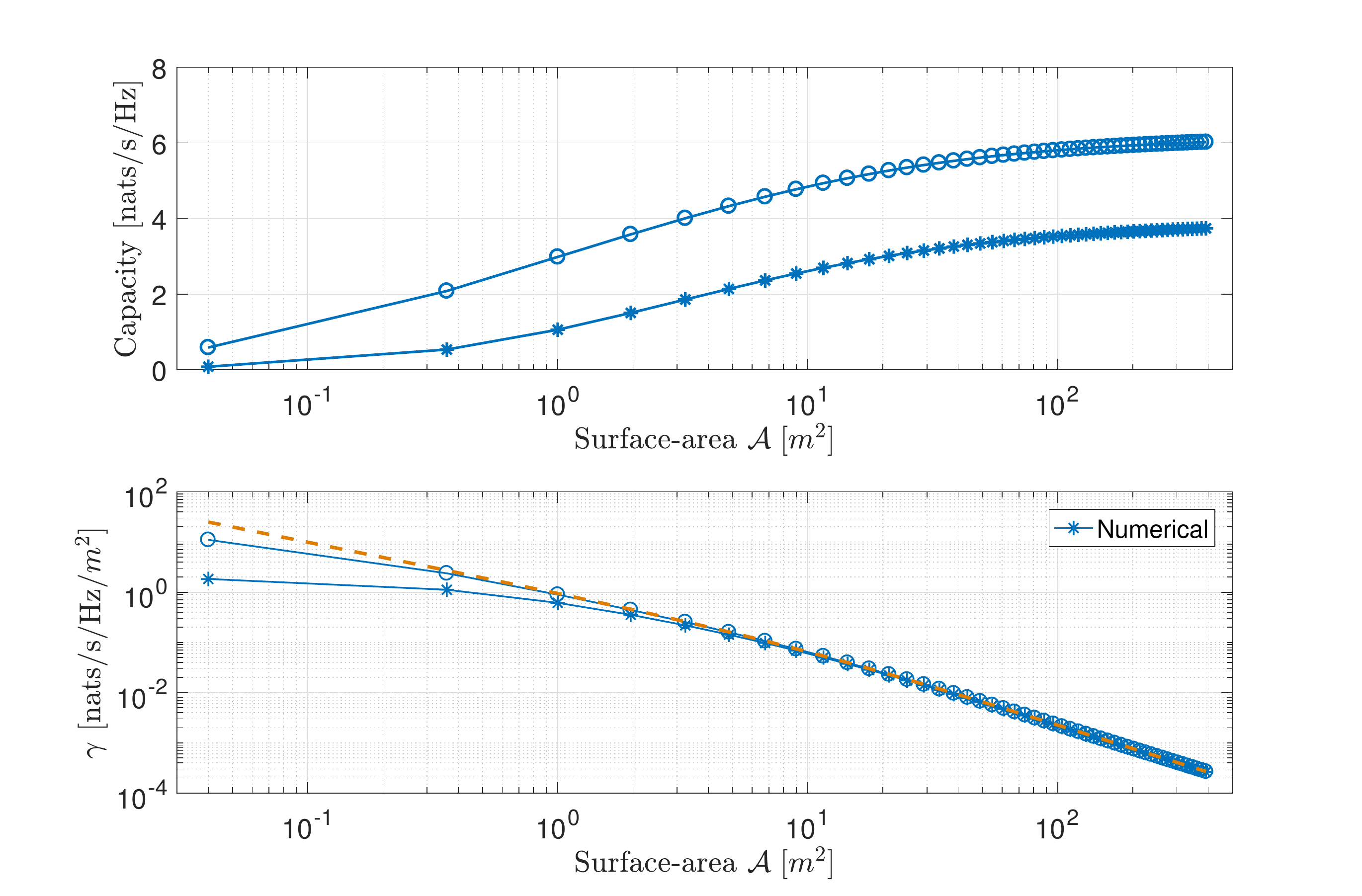}}
\vspace*{-8mm}
\caption{\label{fig2}The capacity $\mathcal{C}$ and utility $\gamma$ of a LIS without HWI, and with settings $z_0\!=\!4$, noise PSD $N_0\!=\!1$. The transmit power is $P\!=\!20$ dB (with marker \lq{}*\rq{}) and 30 dB (with marker \lq{}o\rq{}), respectively. The higher the transmit power is, the closer the utility is to the upper-bound.}
\vspace*{-6mm}
\end{center}
\end{figure}

\subsection{The Decreasing Rate of the Utility}
With the bound $\gamma_0$ in (\ref{ubd}), we can analyze the changes of utility of the surface-area. Firstly, we notice that as $\tau\!\to\!\infty$, $\gamma_0$ decrease to 0, which is natural since when the surface-area is infinitely large the capacity saturates. Secondly, when the surface-area is small, i.e., $\tau\!\to\!0$, the LIS has the highest utility and it holds that
\bea  \gamma_0\approx\frac{1}{4z_0^2}\frac{1}{\arctan\left(\frac{\tau^2}{\sqrt{2\tau^2+1}} \right)}\approx \frac{1}{4\tau^2z_0^2}=\frac{1}{\mathcal{A}}. \notag \eea
Summarizing the discussions above, we can have the below property.
\begin{property}
Without HWI and a small $N_0$ (such that $\gamma$ is close to $\gamma_0$), we have
\bea \label{prop1} \gamma \mathcal{A}\approx 1 \eea
for a small surface-area $\mathcal{A}$ of the LIS.
\end{property}
We plot both the capacity and utility in Fig. 2 for cases without HWI, where we see that the upper-bound of the utility is tight as $\mathcal{A}$ or transmit power $P$ increases. Further, Property 1 can also be clearly seen from Fig. \ref{fig2}, where $\log\gamma +\log\mathcal{A}\approx 0$ holds when the surface-area is small.

\section{Degradations Caused by HWI}

Next we discuss the degradations in capacity and utility of the LIS systems caused by the HWI. 

%

\subsection{General HWI Model}
Without loss of generality, we model the HWI in a general form as
\bea \label{md1} f(r)=\alpha r^{2\beta}, \eea
where $\alpha$ and $\beta$ are non-negative, and when $\alpha\!=\!0$ it holds that $f(r)\!=\!0$ which represent the case without HWI. The model (\ref{md1}) is meaningful in the sense that any other form of $f(r)$ (with similar properties) can be approximated by its Taylor series whose components are of the forms in (\ref{md1}).

\subsection{Capacity Degradation}
With the model (\ref{md1}), the effective noise density in (\ref{tN}) with HWI equals
\be \label{tN1} \tilde{N}=N_0+\frac{Pz_0\alpha }{4\pi}\;\,\frac{\iint\limits_{0\leq x,\,y\leq A}r^{2\beta}\big(x^2+y^2+z_0^2\big)^{-3} \mathrm{d}x\mathrm{d}y}{\iint\limits_{0\leq x,\,y\leq A}\big(x^2+y^2+z_0^2\big)^{-\frac{3}{2}} \mathrm{d}x\mathrm{d}y} . \ee
Analytical analysis of (\ref{tN1}) is possible, but the form of (\ref{tN1}) is rather complicated, and it is hard to obtain insights on the impact of HWI. Alternatively, as shown in Fig. 2, we are more interested in the case when $\tau$ is small which yields a high utility. Therefore, we assume that $A\!\ll\!z_0$, and (\ref{tN1}) can then be simplified into
\bea \label{tN2} \tilde{N}\approx N_0+\frac{P\alpha}{4\pi z_0^2A^2}\iint\limits_{0\leq x,\,y\leq A}\!\!\big(x^2+y^2\big)^{\beta} \mathrm{d}x\mathrm{d}y. 
\eea
To further simplify the integrals, we approximate the square shaped domain $-A\!\leq\! x,\,y\!\leq\! A$ by a disk $x^2\!+\!y^2 \!\leq \!4A^2/\pi$ with the same area. Then, (\ref{tN2}) can be simplified into
\bea \label{tN3} \tilde{N}\!\!\!\!&\approx&\!\!\!\!N_0+\frac{P\alpha}{8 z_0^2 A^2}\int_0^{\frac{2A}{\sqrt{\pi}}}\!\!r^{2\beta+1} \mathrm{d}r \notag \\
\!\!\!\!&=&\!\!\!\!N_0+\frac{4^{\beta-1}P\alpha A^{2\beta}}{(\beta+1)z_0^2\pi^{\beta+1}}. \eea
Note that, the impact of the surface-area is $A^{2\beta}$ (i.e., in the order $\beta$ of the surface-area) in (\ref{tN3}), and the approximation is exact when there is no HWI, i.e., $\alpha\!=\!0$.

Now we can evaluate the capacity degradation with HWI. As seen from the capacity formula (\ref{C1}), the capacity degradation can be measured in terms of the received SNR losses with HWI, which equals 
\bea  \sigma\approx\frac{\tilde{N}}{N_0 }=1+\frac{4^{\beta-1}P\alpha A^{2\beta}}{ (\beta+1)z_0^2\pi^{\beta+1}N_0}. \eea
Under the case that $\beta\!\ll\!1$, it holds that
\bea  \label{sigma} \sigma\approx1+\frac{P\alpha A^{2\beta}}{ 4\pi z_0^2N_0}.    \eea
where $P/( 4\pi z_0^2)$ is the averaged power (total received power divided by the surface-area) reached at each point on the LIS. Further, under the extreme case $\beta\!=\!0$, the approximation in (\ref{sigma}) is exact (under $A\!\ll\!z_0$).

In Fig. 3, we show the received SNR losses $\sigma$ for different values of $\alpha$ and $\beta$, where we can see that the losses are quite significant when the surface-area increases and for large values of $\alpha$ and $\beta$.

\begin{figure}[t]
\begin{center}
\vspace*{-4mm}
\hspace*{-2mm}
\scalebox{0.35}{\includegraphics{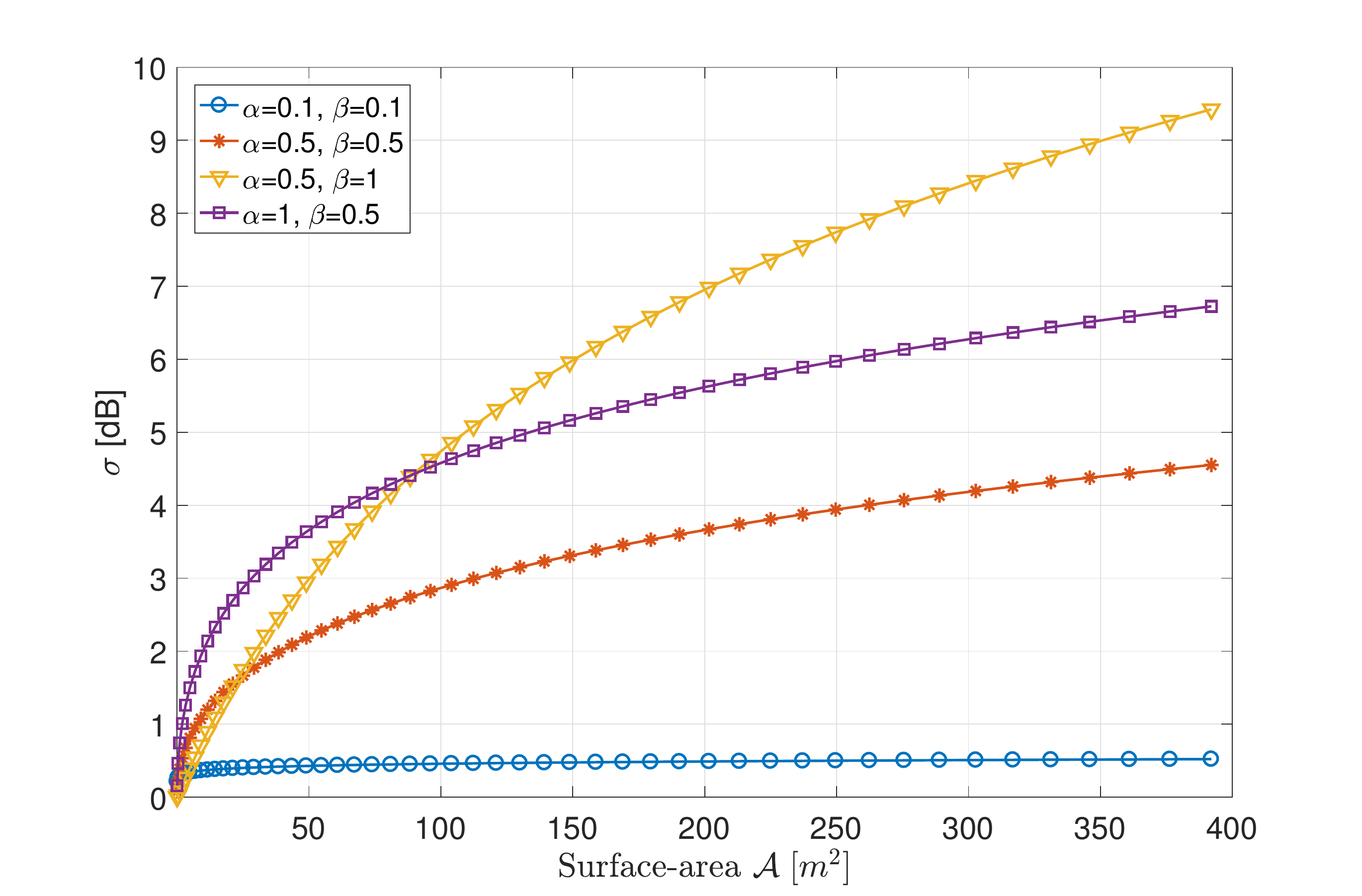}}
\vspace*{-8mm}
\caption{\label{fig3}The degradation in terms of losses of the received SNR ($\sigma$), for different values of $\alpha$ and $\beta$ and under the same settings as Fig.2 with $P\!=\!20$ dB.}
\vspace*{-7mm}
\end{center}
\end{figure}

\subsection{Utility Degradation}
Based on (\ref{tN3}), we can also compute the derivative $\partial \tilde{N}/\partial A$, which equals
\bea \label{dev3}  \frac{\partial \tilde{N}}{\partial A}=\frac{\beta 4^{\beta-\frac{1}{2}}P\alpha A^{2\beta-1}}{ (\beta+1)z_0^2\pi^{\beta+1}}. \eea
From (\ref{gam1}), the degradation of the utility $\gamma$ comprises two aspects. Firstly, the slope of $\gamma$ decreases as 
\bea\frac{P}{8A(N_0+\zeta P)} \longrightarrow \frac{P}{8A(\tilde{N}+\zeta P)}. \eea
Secondly, except for the slope, the utility $\gamma$ is also decreased by an offset
\bea \label{zetaN} \frac{\zeta}{\tilde{N}} \frac{\partial \tilde{N}}{\partial A}=\frac{\zeta\beta 4^{\beta-\frac{1}{2}}P\alpha A^{2\beta-1}}{(\beta+1) z_0^2\pi^{\beta+1}N_0 +4^{\beta-1}P\alpha A^{2\beta}}. \eea

Unlike the case without HWI (where the utility $\gamma$ is always positive and asymptotically decreases to 0 as the surface-area increases), $\gamma$ is negative with HWI whenever
\bea \label{cond1} \frac{\partial \zeta}{\partial A}<\frac{\zeta }{\tilde{N}}\frac{\partial \tilde{N}}{\partial A}. \eea
With equations (\ref{nu}), (\ref{dev1}), and (\ref{zetaN}), the condition (\ref{cond1}) can be rewritten as
\be  \label{cond2}   \frac{1 }{\pi z_0\sqrt{2 \tau^2 \!+\!1} \left(\tau^2\! + \!1\right)} \!<\!\frac{\zeta\beta 4^{\beta-1}P\alpha \tau^{2\beta-2}z_0^{2\beta-3}}{(\beta+1)\pi^{\beta+1}N_0 +4^{\beta-1}P\alpha \tau^{2\beta}z_0^{2\beta-2}}.\ee
That is, when the condition (\ref{cond2}) holds, the utility $\gamma$ is negative and it is not beneficial to increase the surface-area $\mathcal{A}$. Further, when $N_0$ approaches 0, the condition (\ref{cond2}) changes to, after some manipulations,
\bea    \label{cond3}   \beta>\frac{\tau^2 }{ \sqrt{2 \tau^2 +1} \left(\tau^2 + 1\right)\arctan\!\left(\frac{\tau^2}{\sqrt{2\tau^2+1}} \right)}. \eea

Fist of all, we notice that with larger $\alpha$ and $\beta$, the value of the r.h.s gets smaller and the condition (\ref{cond3}) is easier to meet. That is, a negative utility $\gamma$ occurs with severe HWI. Secondly, although the utility $\gamma$ can be negative, when $\tau\!\to\!\infty$, i.e., the surface-area is infinitely large, the utility always converges to zero as $\partial \mathcal{C}/\partial A$ approaches zero. Thirdly, when $\tau\!\to\!0$, i.e., the surface-area is very small, the r.h.s. of (\ref{cond3}) equals 1, and the condition for a negative utility happens is $\beta\!>\!1$ in high SNR cases. While when $\tau\!\to\!\infty$, the r.h.s. of (\ref{cond3}) equals 0. That is, for a sufficiently large LIS with HWI, the utility can always be negative for any $\beta\!>\!0$ (although it converges to zero with an infinitely large surface-area as mentioned above), and the capacity can be decreased if further increasing the surface-area.

Summarizing the discussions above, we have the below Property 2.
\begin{property}
With the general HWI modeled in (\ref{md1}), the capacity with the LIS decreases if the condition (\ref{cond2}) is met. That is, with HWI it is not always beneficial to increase the surface-area.
\end{property}

\begin{figure}[t]
\begin{center}
\vspace*{-4mm}
\hspace*{-5mm}
\scalebox{0.35}{\includegraphics{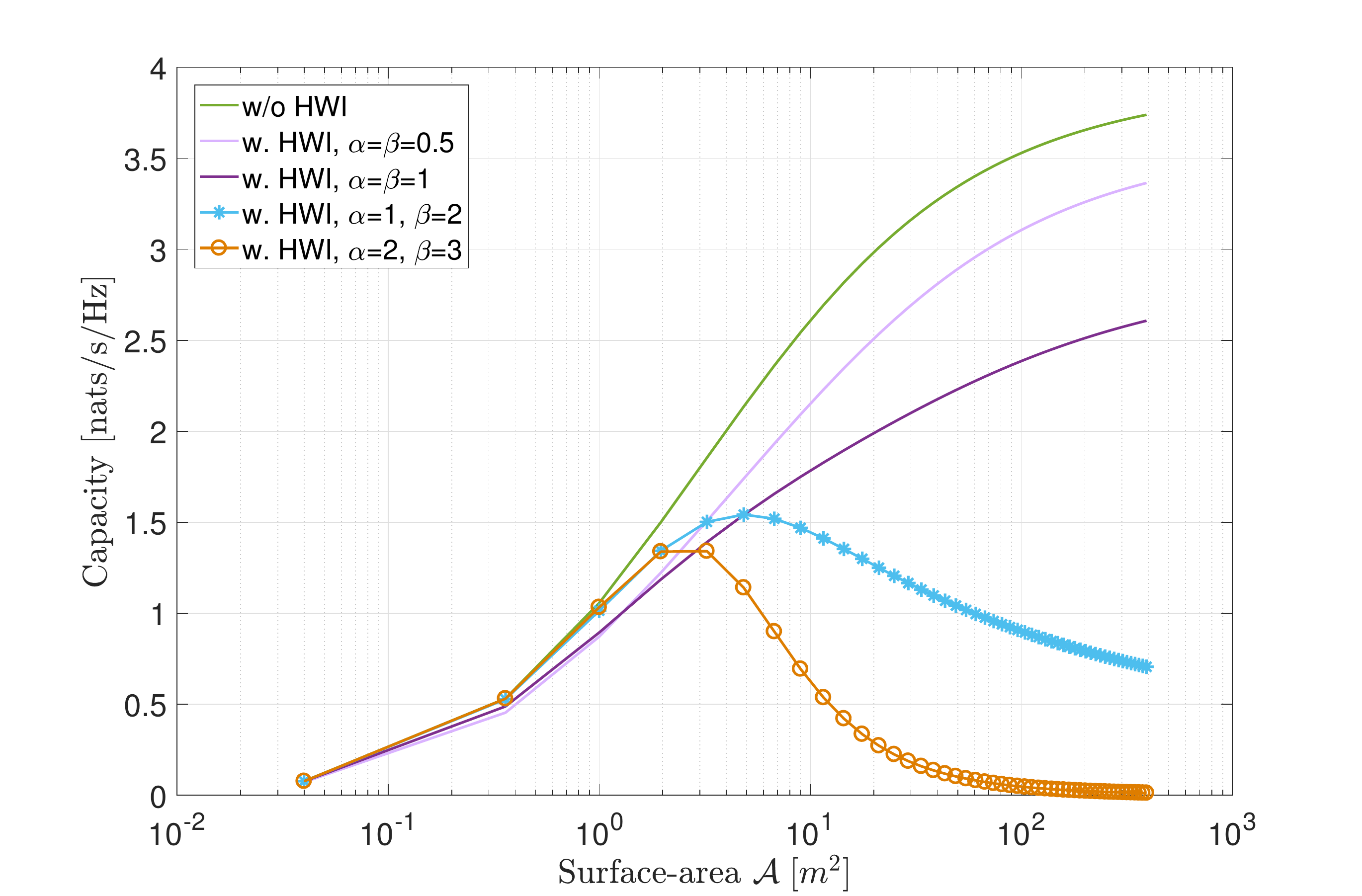}}
\vspace*{-8mm}
\caption{\label{fig4}The capacity degradation under the same settings as Fig.2 with $P\!=\!20$ dB with numerically compute the effective noise density in (\ref{tN1}).}
\vspace*{-5mm}
\end{center}
\end{figure}

\begin{figure}
\begin{center}
\vspace*{-2mm}
\hspace*{-5mm}
\scalebox{0.35}{\includegraphics{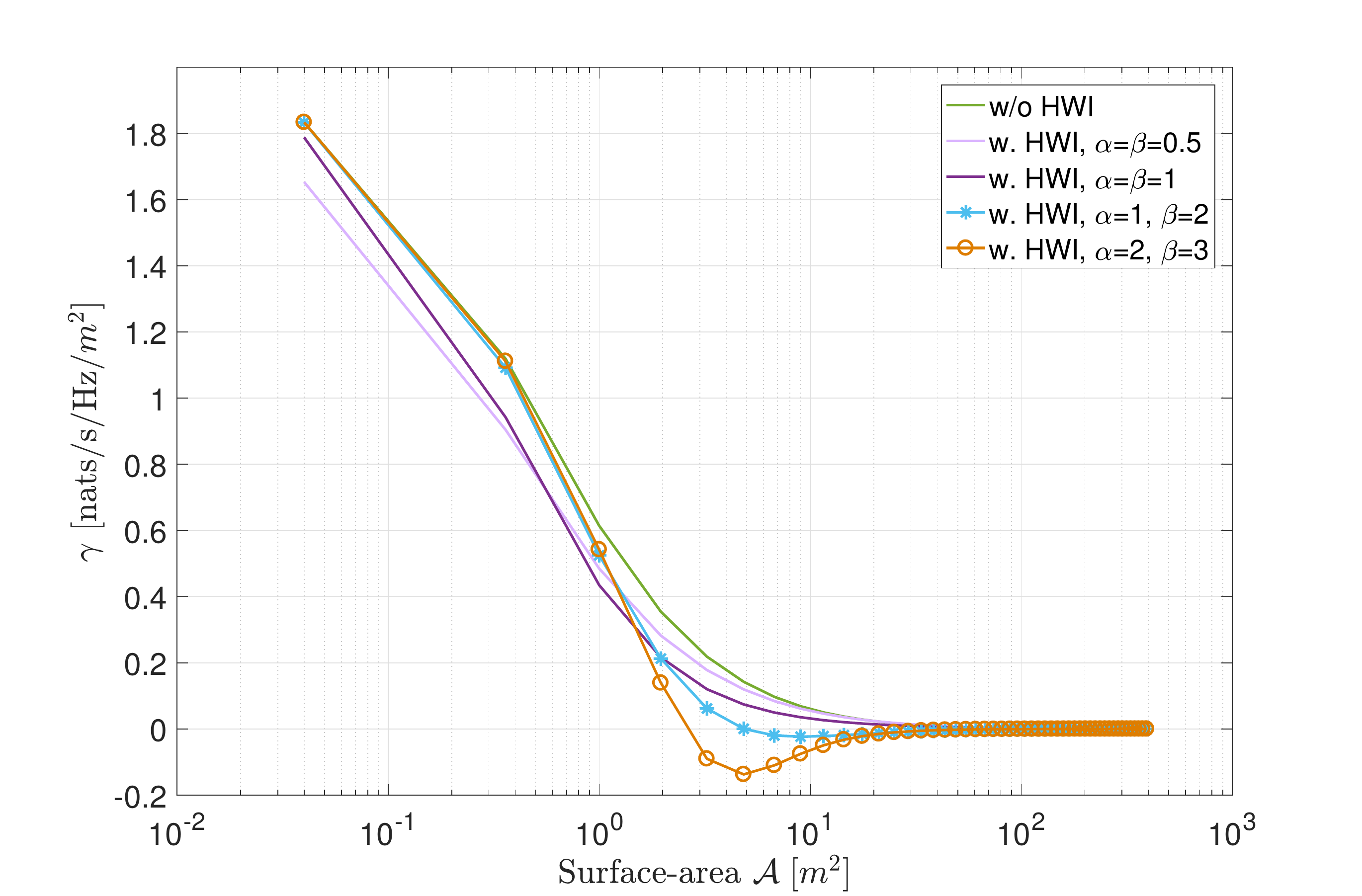}}
\vspace*{-8mm}
\caption{\label{fig5}The utility degradation under the same settings as Fig.4.}
\vspace*{-7mm}
\end{center}
\end{figure}

We show some numerical results in Fig. 4 and 5 to illustrate the capacity and utility with different HWI parameters. As can be seen, the capacity and utility can both be severely decreased with HWI. The larger the values of $\alpha$ and $\beta$ are, the larger the decrements with a sufficiently large surface-area are. More interestingly, according to Property 2 and with conditions (\ref{cond2}) and (\ref{cond3}), the capacity can even decrease with a larger surface-area (with $\beta\!=\!2$ and 3 in the plots). For the case $\alpha\!=\!2$ and $\beta\!=\!3$ (and $P\!=\!20$ dB, $N_0\!=\!1$, and $z_0\!=\!2$ as in Fig. 4), the turning point for $\beta\!=\!0$ according to (\ref{cond2}) is $\tau\!\approx\!0.3827$, where according to the simulation results in Fig. 5, the value $\gamma\!=\!0$ corresponds to a surface-area 2.66 m$^2$, or equivalently, $\tau\!\approx\!0.4077$. Therefore, these two values are close, and the error is due to our approximations ($A\!\ll\!z_0$ and replacing the square shape of the LIS by a disk) in (\ref{tN3}) for deriving the analytical expression.

\section{Splitting the LIS into Multiple Small LIS-Units}
To reduce the impacts of HWI when implementing a LIS system, as can be seen from the discussions in Sec. III, it is beneficial to split the LIS into an array of small LIS-Units. For each LIS-Unit, there is a standalone SPU that collects the signal, with a synthesis unit that combines the signal from all SPUs. As for each LIS-Unit, the surface-area $\mathcal{A}$ is small, the HWI can be reduced. Such a distributed implementation of the LIS system increases the complexity and costs for synchronizing different process units and synthesizing the received signal from different units. Hence, it is of interest to analyze the possible gains with such a distributed deployment as depicted in Fig. 6.

First of all, the received signal power for the distributed deployment remains the same as a single LIS with the same surface-area, which is $\zeta P$. The advantage comes from the decrement of the noise density $\tilde{N}$ by splitting a single LIS into $M$ small LIS-Units, which from (\ref{tN}) is upper-bounded by
\bea  \label{tN4} \tilde{N}\leq N_0+\tilde{N}_s\left(\frac{A}{M}\right)\!, \eea
where $\tilde{N}_s(A)$ denotes the HWI for a LIS-Unit with a length $A$ and and a user located at the CPL of the LIS-Unit. The upper-bound comes from the CPL assumption, that is, the user is located at the CPL for all LIS-Units, which is only approximately true when $A\!\ll\!z_0$. Nevertheless, from (\ref{tN3}) it holds that
\bea  \label{tNs} \tilde{N}_s(A)=\frac{4^{\beta-1}P\alpha A^{2\beta}}{ (\beta+1)z_0^2\pi^{\beta+1}}.\eea
Inserting (\ref{tNs}) back into (\ref{tN4}) yields
\bea  \label{tN5} \tilde{N}\leq N_0+\frac{1}{M^{2\beta}}\frac{4^{\beta-1} P\alpha A^{2\beta}}{ (\beta+1)z_0^2\pi^{\beta+1}}. \eea
That is, the HWI is decreased at the order of $M^{2\beta}$, and under the cases that $\beta\!=\!0$, it is not beneficial to split the LIS into small pieces for HWI suppression.

\begin{figure}[t]
\begin{center}
\vspace*{-4mm}
\hspace*{-7mm}
\scalebox{0.75}{\includegraphics{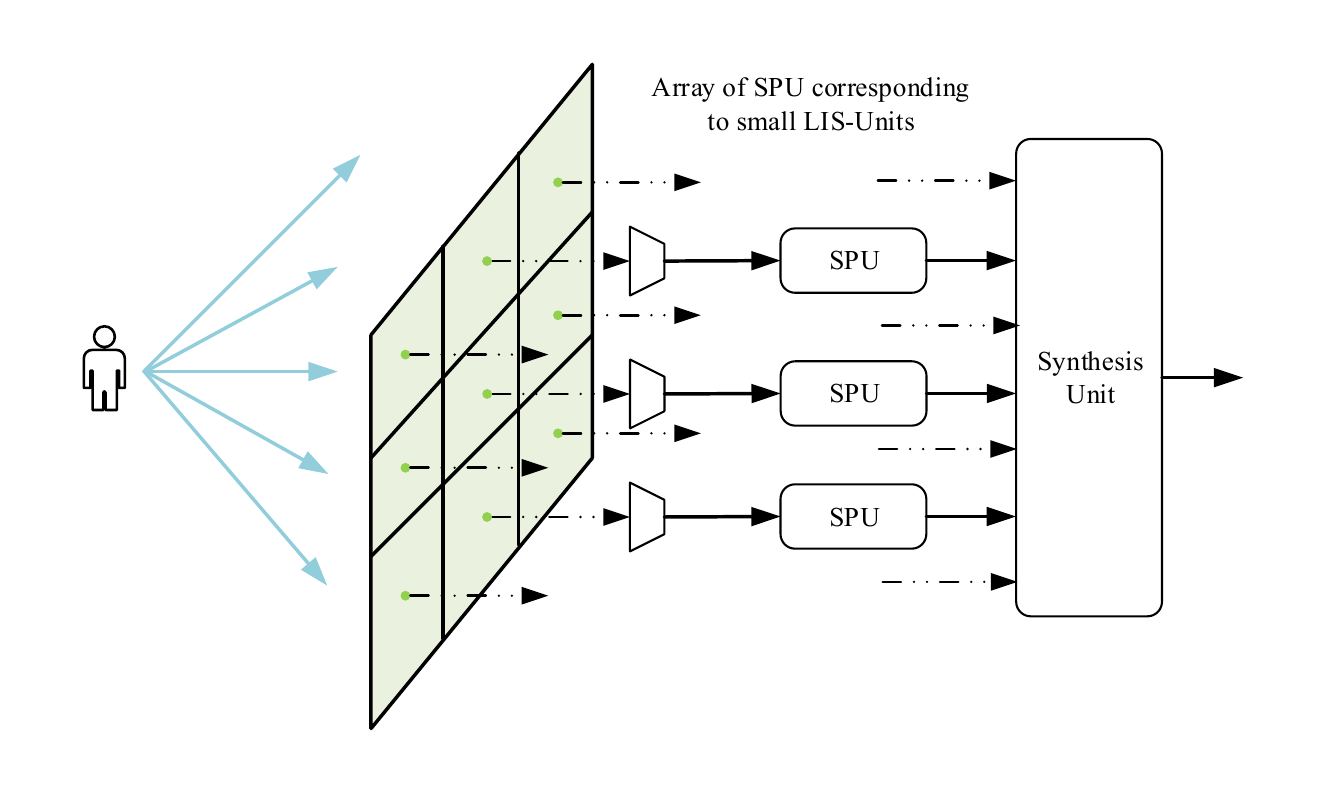}}
\vspace*{-12mm}
\caption{\label{fig7}Splitting a LIS into a number of small LIS-Units to suppress HWI, where each small LIS-Unit has a standalone SPU and the HWI is limited by its small surface-area.}
\vspace*{-5mm}
\end{center}
\end{figure}

\begin{figure}
\begin{center}
\vspace*{-2mm}
\hspace*{-5mm}
\scalebox{0.35}{\includegraphics{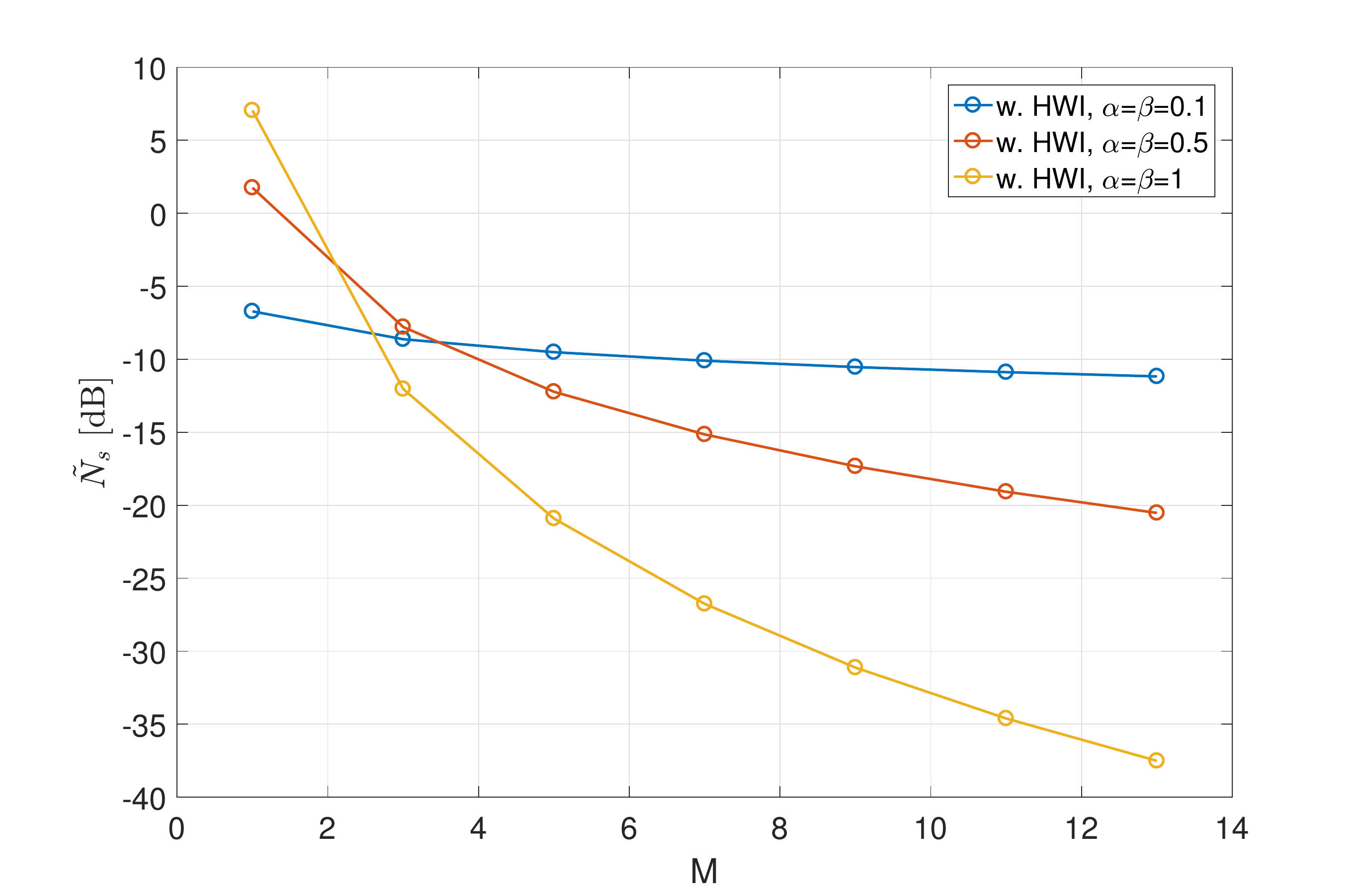}}
\vspace*{-8mm}
\caption{\label{fig8}The decrements of HWI by splitting a single LIS with surface-area $\mathcal{A}\!=\!16$ m$^2$ into $M$ small LIS-Units.}
\vspace*{-7mm}
\end{center}
\end{figure}

In Fig. \ref{fig8}, we show the HWI decrements by splitting a single LIS with surface-area $\mathcal{A}\!=\!16$ m$^2$ into $M$ small LIS-Units. As can be seen, the HWI are significantly decreased with splitting the LIS into 3 or 5 small units. By splitting the LIS into 11 small units, the HWI are almost negligible. Note that, the larger the $\beta$, the smaller $\tilde{N}_s(A)$ it gets when the surface-area is small.

\section{Summary}
We have considered capacity effects of hardware impairments (HWI) in large intelligent surface (LIS) systems. We have modeled the HWI in a general form in relation to the distance from a considered point on the LIS to its center, with the latter one used as a reference-point in hardware designs. Without HWI, the utility of surface-area in terms of capacity is proportional to the inverse of the surface-area for a small LIS. However, with HWI the capacity can be significantly decreased and we have derived closed-form expressions for both the capacity and the utility of surface-area. With severe HWI, increasing surface-area can decease the capacity which should be considered in practical implementations of the LIS system. Further, we have also considered the case with splitting a large LIS into an array comprising a number of small LIS-Units, and we have shown that the HWI can be greatly suppressed with such an approach.

\section*{Appendix: The Proof of (\ref{tN}) }
With applying an MF (with normalization) on (\ref{r1}) and integrating over the entire surface, the received signal equals
\bea \label{r2} \tilde{r}=\frac{1}{\sqrt{\zeta}}\iint\limits_{(x,\,y)\in\mathcal{S}}s^{\ast}(x,y)r(x,y) \mathrm{d}x\mathrm{d}y    =\sqrt{P\zeta} a +\frac{\tilde{w}}{\sqrt{\zeta}}  \notag,\eea
where
\be \tilde{w}\!=\!\iint\limits_{(x,\,y)\in\mathcal{S}}\!\sqrt{P}h(r)|s(x,y)|^2\mathrm{d}x\mathrm{d}y+\iint\limits_{(x,\,y)\in\mathcal{S}}\!s^{\ast}(x,y)n(x,y)\mathrm{d}x\mathrm{d}y . \notag \ee
Since the noise and the HWI are independent and white Gaussian variables, it holds that
\bea \frac{1}{\zeta}\mathbb{E}(\tilde{w}\tilde{w}\rmh)\!\!\!\!&=&\!\!\!\!P\left|\;\,\iint\limits_{(x,\,y)\in\mathcal{S}}h(r)|s(x,y)|^2\mathrm{d}x\mathrm{d}y\right|^2 \notag \\
&&\!\!\!\!+\left|\;\,\iint\limits_{(x,\,y)\in\mathcal{S}}s^{\ast}(x,y)n(x,y)\mathrm{d}x\mathrm{d}y\right|^2  \notag \\
\!\!\!\!&=&\!\!\!\! \frac{P}{\zeta}\iint\limits_{(x,\,y)\in\mathcal{S}}f(r)|s(x,y)|^4\mathrm{d}x\mathrm{d}y +N_0\notag, \eea
which is the noise density in (\ref{tN}), after some manipulations.

\end{document}